\documentclass[12pt]{article}

\pdfoutput=1

\usepackage{hyperref}
\usepackage{graphicx}
\usepackage{longtable}
\usepackage{revsymb}

\def\la{\mathrel{\mathpalette\fun <}}
\def\ga{\mathrel{\mathpalette\fun >}}
\def\fun#1#2{\lower3.6pt\vbox{\baselineskip0pt\lineskip.9pt
\ialign{$\mathsurround=0pt#1\hfil##\hfil$\crcr#2\crcr\sim\crcr}}}

\begin{document}

\title{\bf Critical nucleus charge in a superstrong magnetic
field: effect of screening}
\author{S.I.~Godunov \\
\small{\em ITEP, Moscow} \\ B.~Machet \\
\small{\em LPTHE, UMR 7589 (CNRS \ UPMC Univ Paris 06), Paris} \\
M.~I.~Vysotsky \\
\small{\em ITEP, Moscow}}
\date{}
\maketitle

\begin{abstract}
  A superstrong magnetic field stimulates the spontaneous production of
  positrons by naked nuclei by diminishing the value of the critical
  charge $Z_{cr}$. The phenomenon of screening of the Coulomb potential by
  a superstrong magnetic field which has been discovered recently
  acts in the opposite direction and
  prevents the nuclei with $Z<52$ from becoming critical. For $Z>52$
  for a nucleus to become critical stronger $B$ are needed than
  without taking screening into account.  
\end{abstract}


\maketitle

\section{Introduction}

In a superstrong external magnetic field the Coulomb potential becomes
screened \cite{1,1a}. The screening occurs at the one-loop level; the
corresponding Feynman diagram is the polarization operator
insertion into the photon propagator. This phenomenon leads to the
finiteness of the ground state energy of a hydrogen atom in the
limit of infinite magnetic field $B$ (without screening the ground
state energy diverges as $-\ln^2 B$). In \cite{2a} an analytical
formula for the screened Coulomb potential has been derived. It describes
the behavior of the potential along the magnetic field and
determines the atomic energies. In Sect. II of this paper we will
derive the formula which describes the behavior of the screened
Coulomb potential in the direction transverse to the magnetic
field. We will see that for $B\gg m^2/e^3$ in the transverse
 direction the Coulomb potential is screened at
all distances $\rho \equiv \sqrt{x^{2}+y^{2}} > 1/\sqrt{e^3 B}$ unlike
in the longitudinal
direction, where the screening takes place at $1/m_e >z >
1/\sqrt{e^3 B}$, and in complete analogy with $D=2$ QED with light
fermions analyzed in \cite{2} . In Sect. III we
will investigate how the effects of higher loops modify the one-loop
result for the Coulomb potential. The contributions of higher loops
are suppressed by powers of the fine structure constant
$\alpha$. So with high accuracy the potential
is determined by the one-loop result.

In papers \cite{1,1a,2a} the spectrum of energies on which the lowest Landau
level (LLL) splits in the proton electric field 
was found by solving the corresponding Schr\"{o}dinger equation.
According to \cite{2a} the ground state energy of hydrogen in the
limit of infinite $B$ equals $E_0 = -1.7$ keV, so the use of
the nonrelativistic Schr\"{o}dinger equation is at least
selfconsistent. However,  the size $a_{H}$ of the electron wave function for
$B > m_e^2/e^3$ in the direction transverse to the magnetic field
is much smaller than the electron Compton wavelength,
$a_H\equiv 1/\sqrt{eB} < e/m_e\ll 1/m_{e}$, which makes the nonrelativistic approach a bit
suspicious. That is why starting from Sect IV we will study the ground
state energy of the electron
in a hydrogen-like ion in the presence of an external magnetic field
by analyzing the Dirac equation. Without taking screening into account
this problem was considered in paper \cite{3} (see also \cite{55}),
soon after it was found that a hydrogen-like ion becomes critical at
$Z=170$: the electron ground level sinks into the lower continuum
($\varepsilon_{0}<-m_{e}$) and the vacuum becomes unstable by
spontaneous $e^{+}e^{-}$ pairs  production. These results were
obtained by solving the Dirac equation for an electron moving in the
field of a nucleus of finite  radius. That the phenomenon of
criticality appears only in the framework of the Dirac equation is an
additional motivation to go from Schr\"{o}dinger to Dirac.

According to \cite{3,55} the external magnetic field diminishes the
value of the atomic charge $Z_{cr}$ at which the electron ground level
enters the lower continuum. It happens because a large magnetic field
makes the electron motion quasi-one-dimensional, and in $d=1$ the 
potential $1/|z|$ is more singular than in $d=3$ the potential $1/r$.

In Sect. IV from the numerical solution of the Dirac equation for the
ground electron level of a hydrogen atom in the Coulomb
potential we will find that the corrections to the nonrelativistic
results are small and that the estimate $\delta E \equiv |E_0^{\rm D} -
E_0^{\rm Sch}| \sim (E_0^{\rm Sch})^2/m_e$ works well.

In Sect. V we will study how screening modifies the results of paper
\cite{3}\footnote{The necessity of such consideration was
stressed in \cite{1a}.}. The value of the magnetic field $B_{\rm cr}^Z$
at which an ion with charge $Z$ becomes critical increases because
of screening and only ions with $Z \ga 52$ can become critical.

Let us point out at a major difference between our results and those
of \cite{3}. In \cite{3} the lower limit on $Z$ above which the ion becomes
critical originates from the non-zero size of its radius, which
``cuts off'' the singularity of the Coulomb potential. At the
opposite, our limit $Z\geq 50$ is universal and holds true even for
a point-like nucleus. That the ion has a finite size only slightly
strengthens the constraint.

Our results are summarized in Table \ref{table4} and Fig. \ref{fig:Bcr}, in which the value
of $B/B_{0}$ at which the ion becomes critical is plotted as a
function of its electric charge $Z_{cr}$. The results including
screening are given by the blue curve. If one omits screening, one
gets the dashed-green curve: at the same value of $Z_{cr}$,
criticality with screening is seen to occur at larger B, enormously
larger for $Z=50$. The underlying mechanism that prevents its
occurrence is the freezing of the ground state energy of the
electron, which cannot go into the lower continuum. For $Z>20$ the
screening of the ground state energies occurs in a relativistic regime.

\section{${\bf 3D}$ pattern of the screened Coulomb potential}

Our starting point is the expression for the electric potential in
the momentum representation $\Phi(k)$ where the one-loop contribution to the photon
polarization operator in the external magnetic field is taken into
account. It greatly simplifies
when the external magnetic field is larger than the Schwinger field: $B
> B_0 \equiv m_e^2/e$ (we use Gauss units, $e^2 = \alpha = 1/137.03...$).
For such a strong magnetic field the polarization operator is dominated
by the contribution of electrons which occupy LLL, and a simple
and rather accurate interpolation formula for it was suggested in
\cite{2}. With the help of (8) and (12) from \cite{2} we
obtain:
\begin{equation}
\Phi(k) = \frac{4\pi e}{k_\parallel^2 + k_\bot^2 + \frac{2e^3
B}{\pi} \exp(-\frac{k_\bot^2}{2eB})\frac{k_\parallel^2}{6m_e^2 +
k_\parallel^2}} \;\; , \label{1}
\end{equation}
and
\begin{equation}
\Phi(z,\rho) = 4\pi e\int\frac{e^{i\bar{k}_\bot \bar{\rho} +
ik_\parallel z} d k_\parallel d^2 k_\bot/(2\pi)^3}{k_\parallel^2 +
k_\bot^2 + \frac{2e^3 B}{\pi}
\exp(-\frac{k_\bot^2}{2eB})\frac{k_\parallel^2}{6m_e^2 +
k_\parallel^2}} \;\; , \label{2}
\end{equation}
where we assume that the magnetic field $B$ is directed along the $z$ axis
and $k_{\parallel}$ is the momentum component parallel to $B$, while
$\bar{k}_{\bot}$ and $\bar{\rho}$ are vectors in the plane transverse to
the magnetic field.

The expression for the electric potential of a pointlike charge in
the direction of the magnetic field at $\rho =0$ was obtained in
\cite{2a}:
\begin{equation}
{\bf\Phi}(z,0) = \frac{e}{|z|}\left[ 1-e^{-\sqrt{6m_e^2}|z|} +
e^{-\sqrt{(2/\pi) e^3 B + 6m_e^2}|z|}\right] \;\; . \label{3}
\end{equation}

For $B\ll 3\pi m_e^2 /e^3$ the potential equals Coulomb up to small,
power suppressed, terms, while for $B\gg 3\pi m_e^2/e^3$ we get:
\begin{equation}
{\bf\Phi}(z,0) = \left\{
\begin{array}{lll}
\frac{e}{|z|} e^{-\sqrt{(2/\pi) e^3 B}|z|}          & , & |z| < l_{0} \\
\frac{e}{|z|}\left(1- e^{-\sqrt{6m_e^2}|z|}\right) & , & l_{0} < |z| < \frac{1}{m_e}
 \\
\frac{e}{|z|} & , & \frac{1}{m_e} < |z|
\end{array}
\right. \;\; , \label{4}
\end{equation}
where $l_{0}=\frac{1}{\sqrt{(2/\pi) e^3 B}}\ln\left(\sqrt{\frac{e^3 B}{3\pi
m_e^2}}\right)$.

The behavior of the potential in the transverse plane ($z=0$) can also be
found analytically in the limit $B\gg 3\pi m_e^2/e^3$. Performing
the integration in (\ref{2}) (for $\rho \ga a_H$
the exponent in the denominator can be neglected) we obtain:
\begin{eqnarray}
\Phi(0,\rho)  = \left\{
\begin{array}{lll}
\frac{e}{\rho}\exp(-\sqrt{(2/\pi)e^3 B}\rho)& , & \rho < l_{0}
\\
\sqrt{\frac{3\pi m_e^2}{e^3 B}}\frac{e}{\rho}& , & l_{0} < \rho \;\; ,
\end{array}
\right. \label{5}
\end{eqnarray}
and the Coulomb potential is screened at large $\rho$ in complete
analogy with the $D=2$ case, see \cite{2}, Eq. (10).

For $|z| \gg 1/m_e$ the values $|k_\parallel| \ll m_e$ dominate in the
integral (\ref{2}) and we get: 
\begin{equation}
\Phi(\rho, z)\left|_{z\gg 1/m_e}\right. = \frac{e}{\sqrt{z^2 +
(1+\frac{e^3 B}{3\pi m_e^2})\rho^2}} \;\; . \label{6}
\end{equation}

\begin{figure}[t]
  \begin{center}
    \includegraphics[scale=0.8]{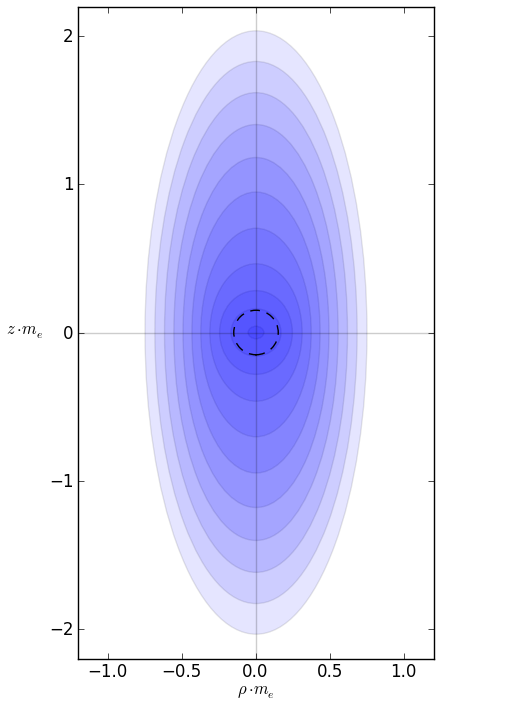}
  \end{center}
  \caption{\label{fig:1}The equipotential lines at $B/B_{0}=10^{4}$. The dashed line
    corresponds to $\sqrt{z^{2}+\rho^{2}}=\frac{1}{\sqrt{(2/\pi) e^3
        B}}\ln\sqrt{\frac{e^3 B}{3\pi m_e^2}}.$} 
\end{figure}

In Fig. \ref{fig:1} the equipotential lines are shown. The behavior of the screened
Coulomb potential in the transverse plane was found numerically in
\cite{1a}, Fig. 2. $\Phi(0,\rho)$ given by expression (\ref{5})
is close to the results shown in Fig. 2 in \cite{1a} for $b \equiv B/B_0 =
10^4 \div 10^6$, within the domain of validity of (\ref{5}), and
deviate from the curve for $b=10^3$ since it
corresponds to $B < 3\pi m_e^2/e^3$, where (\ref{5}) is
inapplicable.

Finally for $3\pi m^2/e^3 \gg B \gg m^2/e$ expanding (\ref{6}) we
get ($\theta$ is the angle between the two vectors $\bar r$ and
$\bar B$):
\begin{eqnarray}
\Phi(r) = \frac{e}{r}\left\{1-\frac{\alpha}{6\pi}(B/B_{\rm
0})\sin^2\theta\right\}\; , \;\; \;\;  r \equiv \sqrt{\rho^2 + z^2} \gg 1/m_{e}
\;\; , \label{7} 
\end{eqnarray}
which coincides with the result obtained in \cite{4} where the
expression for the photon polarization operator at $B > B_0$ was
obtained as well.

\section{Higher loops}

The expression for the screened Coulomb potential was obtained in
\cite{1}-\cite{2a} from the one-loop contribution to the photon
polarization operator in an external magnetic field $B\gg B_0$. In
momentum space it looks like (see (13) in \cite{2a}):
\begin{equation}
\Phi(k_\parallel, k_0 = k_\bot = 0) = \frac{4\pi e}{k_\parallel^2
+ \frac{2e^3 B}{\pi}\frac{k_\parallel^2}{k_\parallel^2 + 6m_e^2}}
\;\; . \label{88}
\end{equation}

If $n$-loop diagrams contain  terms $\sim e^3 B(e^3
B/k_\parallel^2)^{n-1}$ they would drastically change the shape of
the potential in coordinate space.

To calculate the radiative corrections one should use the electron
propagator $G(k)$ in an external homogeneous magnetic field $B$.
Its spectral representation is a sum over Landau levels and for $B \gg
B_0$ the contribution of the lowest level dominates \cite{4, 71}:
\begin{equation}
G(k) = e^{-k_\bot^2/eB}(1-i\gamma_1 \gamma_2)\frac{\hat k_{0,3}
+ m_e}{k_{0,3}^2 - m_e^2} \;\; , \label{89}
\end{equation}
where the magnetic field is directed along the third (or $z$) axis
($\bar{B}=(0,0,B)$), $\hat{k}_{0,3}=k_{0}\gamma_{0}-k_{3}\gamma_{3}$,
$k_{\perp}$ is the component of the momentum normal to the magnetic
field and the projector $(1-i \gamma_1 \gamma_2)$ selects the virtual
electron state with spin opposite to the direction of the magnetic 
field. The contributions of the excited Landau levels
to $G$ yield a term in the denominator proportional to $eB$ and they
produce a correction of order $e^2 \equiv \alpha$ in the denominator
of (\ref{88}). 

Two kind of terms contribute to the polarization operator at the
two-loop level. First, there are terms in the electron propagators
which represent the contributions of higher Landau levels. Just like
in the one-loop case they produce corrections suppressed by $e^2$ in
the denominator of (\ref{88}), i.e. terms of the order $e^5 B$
which can be safely neglected in comparison with the leading $\sim
e^3 B$ term. Second, there is the contribution from the leading term
(\ref{89}) of the electron propagator. Let us consider the
simplest diagram: the photon dressing of the electron propagator.
Neglecting the electron mass we get:
\begin{eqnarray}
\gamma_\mu(1-i\gamma_1 \gamma_2)\hat k_{0,3}\gamma_\mu = -2[\hat
k_{0,3} - i\hat k_{0,3}\gamma_2\gamma_1]=  -2\hat
k_{0,3}(1+i\gamma_1 \gamma_2) \;\; , \label{90}
\end{eqnarray}
which gives zero when multiplied by the external propagator (\ref{89})
of the electron, 
since $(1+i\gamma_1 \gamma_2)(1-i\gamma_1 \gamma_2)=0$. This
result is a manifestation of the following well-known fact: in
$D=2$ massless QED (Schwinger model) all loop diagrams are zero
except the one-loop term in the photon polarization operator. That is
why the two loop diagrams
in which the propagators of virtual electrons are given by (\ref{89}) give
contributions to the polarization operator proportional to $m_{e}^{2}$ for
$k_{\parallel}^{2}\gg m_{e}^{2}$ while in the opposite limit
$k_{\parallel}^{2}\ll m_{e}^{2}$ they should be proportional to
$k_{\parallel}^{2}$ in order for the photon to stay massless. So these
terms are of the order of $\alpha 
(e^{3}B)
\left(m^{2}_{e}k_{\parallel}^{2}/(k_{\parallel}^{2}+m_{e}^{2})^{2}\right)$ and 
they are not important.

The generalization of the above arguments to higher loops is
straightforward.

To conclude this section let us note that an analogous statement about
the unimportance of the two-loop terms was made in \cite{72}. 

\section{Dirac equation with a screened Coulomb potential, ${\bf Z=1}$}

The ground state electron energy of a hydrogen-like ion with electric
charge $Z$ in an external magnetic field was analyzed in \cite{3} in
the framework of the Dirac equation.\footnote{Let us note 
papers \cite{33} in which the relativistic corrections to the hydrogen
spectrum in a strong magnetic field are discussed.}
In a strong magnetic field ($a_H \equiv 1/\sqrt{eB} \ll
1/(m Z e^2))$ the electron spectrum consists of the Landau levels
splitted into Coulomb sublevels; the ground level belongs to LLL. In
complete analogy with the nonrelativistic problem the adiabatic
approximation is applicable. Averaging over the fast motion of the
electron in the plane transverse to the magnetic field,
the Dirac equation for the electron on LLL was reduced in \cite{3} to
two first order one-dimensional differential equations:
\begin{equation}
  \label{eq:11}
  \begin{array}{c}
  g_{z}-(\varepsilon+m_{e}-\bar{V})f=0,\\  
  f_{z}+(\varepsilon-m_{e}-\bar{V})g=0,
\end{array}
\end{equation}
where $\varepsilon$ is the energy eigenvalue of the Dirac equation;
$g_{z}=dg/dz,~f_{z}=df/dz$; the bispinor
$\psi_{e}=\left(\varphi_{e}\atop\chi_{e}\right)$ of the electron is
decomposed into $\varphi_{e}=\left(0\atop
  g(z)\exp\left(-\rho^{2}/4a_{H}^{2}\right)\right)$,
$\chi_{e}=\left(0\atop
  if(z)\exp\left(-\rho^{2}/4a_{H}^{2}\right)\right)$. 

They describe the electron motion in the effective
potential $\bar V(z)$:
\begin{equation}
\bar V(z) = \frac{1}{a_H^2}\int\limits_0^\infty V(\sqrt{\rho^2 +
z^2})\exp\left(-\frac{\rho^2}{2a_H^2}\right)\rho d\rho \;\; , \label{8}
\end{equation}
where $V(r)=-Z e^2/r$, $r^2 \equiv \rho^2 + z^2$. At large
distances $|z|\gg a_H$ the effective potential equals Coulomb, and the
solutions of the equations (\ref{eq:11}) exponentially decreasing at
$|z|\to\infty$ are linear combinations of Whittaker functions. At
short distances the equations (\ref{eq:11}) can be easily integrated for
$|\bar V(z)| \gg |\varepsilon \pm m_e|$, which is equivalent to
the following inequality: $z \ll Z e^2/(2m_e)$. Matching short
and large distance solutions at
\begin{equation}
  \label{eq:8a}
Z e^2/(2m_e)\gg z \gg a_H
\end{equation}
gives an algebraic equation for the ground state energy (it coincides with
Eq. (22) in \cite{3} in the limit $R/a_{H}\ll 1$, where $R$ is the nucleus
radius):
\begin{eqnarray}
 Ze^2\ln\left(2\frac{\sqrt{m_e^2-\varepsilon^2}}{\sqrt{eB}}\right) +
\arctan\left(\sqrt{\frac{m_e+\varepsilon}{m_e-\varepsilon}}\right) +
\arg\Gamma\left(-\frac{Ze^2 \varepsilon}{\sqrt{m_e^2 -
\varepsilon^2}} + iZe^2\right) -\nonumber\\-\arg \Gamma(1+2iZe^2) -
 \frac{Ze^2}{2}(\ln 2 + \gamma) =\frac{\pi}{2}+ n\pi \;\; , \label{9}
\end{eqnarray}
where $\gamma = 0.5772...$ is the Euler constant, and the argument of the
gamma function is given by 
\begin{equation}
\arg\Gamma(x+iy) = -\gamma y + \sum_{k=1}^\infty\left(\frac{y}{k}
- \arctan\frac{y}{x+k-1}\right) \;\; . \label{10}
\end{equation}
For the ground level at $\varepsilon > 0$ one should take $n=0$, while
for $\varepsilon < 0$ it should be changed to $n=-1$.

According to (\ref{9}) when the magnetic field increases the
ground state energy goes down and reaches the lower continuum. The
value of the magnetic field at which this happens is determined by
(\ref{eq:ORS32}) (see below). 

A matching point exists only if $B \gg 4m_e^2/(e(Ze^2)^2)$ (see
(\ref{eq:8a})) and (\ref{9}) is valid only for 
these values of the magnetic field. However, as was
checked in \cite{3} from (\ref{9}) in the non-relativistic regime $Ze^2
\ll 1$, $m-\varepsilon \ll m$, a formula can be deduced which is a
valid solution of the non-relativistic problem and extends the domain
of validity of eq. (\ref{9}).

Thus, without taking screening into account, from (\ref{9}) we can
obtain the dependence of the ground state energy of a hydrogen atom
on the magnetic field for $B \gg 4m_e^2/e^5$. In order to find the
ground state energy at $B \la 4m_e^2/e^5$ and to take screening
into account we solve the equations (\ref{eq:11})
numerically. This system  can be transformed into one second order
differential equation for $g(z)$. By substituting 
$g(z)=\left(\varepsilon+m_{e}-\bar{V}\right)^{1/2}\chi(z)$ 
a Schr\"{o}dinger-like equation for the function $\chi(z)$ was obtained
in \cite{3}:\footnote{This trick was exploited by V.S. Popov for
the qualitative analysis of the phenomenon of critical charge.} 
\begin{equation}
\frac{d^2\chi}{dz^2} + 2m_{e}(E-U)\chi = 0 \;\; , \label{11}
\end{equation}

\begin{eqnarray*}
E = \frac{\varepsilon^2 - m_e^2}{2m_e} \; , \;\;
U = \frac{\varepsilon}{m_e}\bar V- \frac{1}{2m_e}\bar V^2 +
\frac{\bar V''}{4m_e(\varepsilon + m _e-\bar V)} + \frac{3/8(\bar
V')^2}{m_e(\varepsilon +m_e - \bar V)^2} \; ,
\end{eqnarray*}
where $\varepsilon$ is the energy eigenvalue of the Dirac equation and
$\bar V(z)$ is given in (\ref{8}). We integrated (\ref{11}) 
numerically in the present work. Leaving a detailed discussion for a future publication
\cite{6} let us only note that, while for $z \gg 1/m_e$ the last three
terms in the expression for $U$ are much smaller than the first
one (the only one remaining in the nonrelativistic approximation), at
$z \la 1/m_e$ the relativistic terms dominate and are very big for
$B\gg B_0$ at $z \sim a_H$ which makes numerical calculations very
complicated.

\begin{table}[h!]
\caption{\label{tab:1}Values of $\lambda$ for $Z=1$ without
  screening obtained from the Schr\"{o}dinger and Dirac
    equations. They start to differ substantially at enormous values
    of the magnetic field.}
\begin{center}
\begin{tabular}{||c||c|c|c|c||}
\hline
$B/B_{0}$ & KP-equation      & Numerical results   & Eq. (\ref{9}) & Numerical results \\
${~}$    & (Schr\"{o}dinger) &  (Schr\"{o}dinger)   & (Dirac)           & (Dirac)          \\
\hline
$10^{0}$  & 5.737     & 5.735   & 5.735   & 5.734          \\
$10^{1}$  & 7.374     & 7.374   & 7.370   & 7.371          \\
$10^{2}$  & 9.141     & 9.141   & 9.136   & 9.135          \\
$10^{3}$  & 11.00     & 11.00   & 10.99   & 10.99          \\
$10^{4}$  & 12.93     & 12.93   & 12.91   & 12.91          \\
$10^{5}$  & 14.91     & 14.91   & 14.88   & 14.88          \\
$10^{6}$  & 16.93     & 16.93   & 16.89   & 16.89          \\
$10^{7}$  & 18.98     & 18.98   & 18.93   & 18.92          \\
$10^{8}$  & 21.06     & 21.05   & 20.98   & 20.98          \\
$10^{9}$  & 23.16     & 23.15   & 23.05   & 23.05          \\
$10^{10}$ & 25.27     & 25.27   & 25.14   & 25.13          \\
$10^{11}$ & 27.40     & 27.40   & 27.23   & ${~}$          \\
$10^{12}$ & 29.54     & 29.54   & 29.33   & ${~}$          \\
\dots    & \dots     & ${~}$   & \dots   & ${~}$          \\
$10^{15}$ & 36.03     & ${~}$   & 35.64   & ${~}$          \\
\dots    & \dots     & ${~}$   & \dots   & ${~}$          \\
$10^{20}$ & 46.99     & ${~}$   & 46.11   & ${~}$          \\
\dots    & \dots     & ${~}$   & \dots   & ${~}$          \\
$10^{25}$ & 58.07     & ${~}$   & 56.40   & ${~}$          \\
\dots    & \dots     & ${~}$   & \dots   & ${~}$          \\
$10^{30}$ & 69.22     & ${~}$   & 66.38   & ${~}$          \\
\dots    & \dots     & ${~}$   & \dots   & ${~}$          \\
$10^{35}$ & 80.43     & ${~}$   & 75.98   & ${~}$          \\
\dots    & \dots     & ${~}$   & \dots   & ${~}$          \\
$10^{40}$ & 91.67     & ${~}$   & 85.10   & ${~}$          \\
\dots    & \dots     & ${~}$   & \dots   & ${~}$          \\
$10^{45}$ & 102.95    & ${~}$   & 93.67   & ${~}$          \\
\dots    & \dots     & ${~}$   & \dots   & ${~}$          \\
$10^{50}$ & 114.25    & ${~}$   & 101.62   & ${~}$          \\
\dots    & \dots     & ${~}$   & \dots   & ${~}$          \\
$10^{55}$ & 125.57    & ${~}$   & 108.89   & ${~}$          \\
\hline
\end{tabular}
\end{center}
\end{table}

In Table \ref{tab:1} the results for the ground state energy of a hydrogen atom 
without screening are presented. The values of the magnetic field in
units of $B_0$ are given in the first column, while in columns
2-5 the values of $\lambda$ are given. By definition
\footnote{Let us note that the
definition of $\lambda$ used in \cite{3} differs from our:
$\lambda^{\mbox{\cite{3}}} \equiv e^2 \lambda$.}
\begin{equation}
E= \frac{\varepsilon^2 - m_e^2}{2m_e} \equiv -\frac{m_e e^4}{2}\lambda^2 \;\; . \;\;
\label{12}
\end{equation}

From Table \ref{tab:1} we see that:
\begin{enumerate}
\item  the results of the numerical integrations of the
  Schr\"{o}dinger and Dirac equations coincide within four
  digits:
  \begin{itemize}
  \item with the analytical Karnakov--Popov formula for the ground
    state energy ($n_\rho = m = 0$) \cite{7,2a} in the case of the
    Schr\"{o}dinger equation;
  \item with formula (\ref{9}) for $Z = 1$ in the case of the Dirac
    equation;
  \end{itemize}
\item for the relativistic shift of energy the following estimate
  works:
\end{enumerate}
\begin{equation}
E_{\rm Dirac} - E_{\rm Schr} \sim E_{\rm Schr} \frac{E_{\rm
Schr}}{m_e} \;\; , \label{13}
\end{equation}
$$
\delta\lambda \sim e^4 \lambda^3/4 \;\; .
$$

To take screening into account, the following
formula for the effective potential should be used in (\ref{11}) instead
of (\ref{8}):
\begin{eqnarray}
\bar V(z) = -\frac{Ze^2}{a_H^2}\left[1-e^{-\sqrt{6m_e^2}|z|} +
e^{-\sqrt{(2/\pi)e^3 B + 6m_e^2}|z|}\right]\int\limits_0^\infty
\frac{e^{-\rho^2/2a_H^2}}{\sqrt{\rho^2 + z^2}}\rho d \rho \;\; ,\qquad
\label{14}
\end{eqnarray}
where $Z=1$ for hydrogen.

The freezing of the ground state energy is due to a weaker singularity
of the potential with screening (\ref{14}) at $z\rightarrow 0$ for
$B\rightarrow \infty$ than that of the potential without screening
(\ref{8}). While the non-screened potential behaves like $1/z$ at small
$z$, the screened potential is proportional to $\delta(z)$ because,
when $B\rightarrow \infty$, the width of the region where it behaves
like $1/z$ shrinks to zero \cite{1,1a}.

In Table \ref{tab:2} the results of the analytical
formula for $\lambda$ with the account of screening derived in
\cite{2a} for the Schr\"{o}dinger equation are compared with the
results of the numerical integration of the Dirac equation. We see that in
the case of screening the relativistic shift of energy is also
very small, and due to it the ground state energies become a little
bit higher, just like without taking screening into account.
The freezing of the ground state energy occurs at $B/B_0 = 10^3 \div
10^4$, when $B \approx 3\pi m_e^2/e^3$.

\begin{table}[h!]
\caption{\label{tab:2}Values of $\lambda$ for $Z=1$ with screening.}
\begin{center}
\begin{tabular}{||c||c|c|c||}
\hline
$B/B_{0}$   & Eq. (57) from \cite{2a} & Numerical results   &  Numerical results \\
${~}$ &       (Schr\"{o}dinger)             &  (Schr\"{o}dinger)  &   (Dirac)          \\
\hline
$10^{0}$  & 5.7     & 5.7   & 5.7       \\
$10^{1}$  & 7.4     & 7.4   & 7.4       \\
$10^{2}$  & 9.1     & 9.1   & 9.1       \\
$10^{3}$  & 10.5    & 10.6  & 10.6      \\
$10^{4}$  & 11.1    & 11.2  & 11.2      \\
$10^{5}$  & 11.2    & 11.3  & 11.3      \\
$10^{6}$  & 11.2    & 11.4  & 11.3      \\
$10^{7}$  & 11.2    & 11.4  & 11.3      \\
$10^{8}$  & 11.2    & 11.4  & 11.3      \\
\hline
\end{tabular}
\end{center}
\end{table}

\section{Screening versus critical nucleus charge}
\begin{table}[h!]
\caption{\label{tab:3}Values of $\varepsilon_{0}/m_{e}$ for $Z=40$.}
\begin{center}
\begin{tabular}{||c||c|c|c||}
\hline
$B/B_{0}$& Eq. (\ref{9}) & Numerical results   &  Numerical results \\
${~}$   &       (Dirac)      &  (Dirac)            &  with screening (Dirac)\\
\hline
$10^{0}$        & 0.819     & 0.850   & 0.850       \\
$10^{1}$        & 0.653     & 0.667   & 0.667       \\
$10^{2}$        & 0.336     & 0.339   & 0.346       \\
$10^{3}$        &-0.158     &-0.159   &-0.0765      \\
$10^{4}$        &-0.758     &-0.759   &-0.376       \\
$2\cdot 10^{4}$ &-0.926     &-0.927   &-0.423       \\
\cline{2-3}
\dots          &\multicolumn{2}{c|}{at $B/B_{0}\approx 2.85\cdot 10^{4},~ \varepsilon_{0}=-m_{e}$} & \dots \\
\cline{2-3}
$10^{5}$        & ---       & ---     &-0.488       \\
$10^{6}$        & ---       & ---     &-0.524       \\
$10^{7}$        & ---       & ---     &-0.535       \\
$10^{8}$        & ---       & ---     &-0.538       \\
\hline
\end{tabular}
\end{center}
\end{table}

According to \cite{3} nuclei with $Z\geq 40$ become critical in
an external $B$ (for smaller $Z$ the values of $a_H$ at which the
criticality is reached become smaller than the nucleus radius, the Coulomb
potential diminishes and thus the ground level does not reach the lower
continuum). 
\begin{table}[h!]
\caption{\label{tab:freezing_energy}Values of freezing ground state energies for different $Z$
    from the Schr\"{o}dinger and the Dirac equations. In order to find
  the freezing energies we must take $B/B_{0}\gg 3\pi/e^{2}$. In numerical calculations we took
  $B/B_{0}=10^{8}$.}
\centering
\begin{tabular}{||c||c|c||}
  \hline
  $Z$ & 
  $\left(E_{0}^{fr}\right)^{numerical}_{Schr}$, keV &
  $\left(\varepsilon_{0}^{fr}-m_{e}\right)^{numerical}_{Dirac}$, keV \\
  \hline
    1    &     -1.7  & -1.7 \\
    10   &     -88   & -87  \\
    20   &     -288  & -273 \\
    30   &     -582  & -519 \\
    40   &     -966  & -787 \\    
    49   &     -     & -1003\\    
  \hline
\end{tabular}
\end{table}

In Table \ref{tab:3} one can
see the dependence of the ground state electron energy $\varepsilon_{0}$
on the external magnetic field for $Z=40$. The numerical solutions of
(\ref{11}) are in good correspondence with the values of
$\varepsilon_{0}$ obtained from (\ref{9}). The numerical results with
screening are shown in the last column; we see that freezing
occurs in the relativistic domain $\varepsilon_0 \approx -m_e/2$
and the ground level never reaches lower continuum, $\varepsilon_0
> -m_{e}$.

In Table \ref{tab:freezing_energy}  we compare freezing energies for
different $Z$ obtained numerically from the nonrelativistic
Schr\"{o}dinger equation and from the Dirac equation. We see that for
$Z > 20$ the freezing occurs in the relativistic regime, where the
Schr\"{o}dinger equation should not be used. Let us stress that
  the value of $B$ at which the freezing occurs does not depend on $Z$.

From (\ref{9}) we obtain in the limiting case $\varepsilon\rightarrow
-m_{e}$ an equation which defines the value of the magnetic field at
which a nucleus with charge $Z$ becomes critical without taking screening
into account (it coincides with Eq. (32) from \cite{3}):
\begin{equation}
  \label{eq:ORS32}
  \frac{B}{B_{0}}=2(Z_{cr}e^{2})^{2}\exp\left(-\gamma+\frac{\pi-2\arg\Gamma
        (1+2iZ_{cr}e^{2})}{Z_{cr}e^{2}}\right).
\end{equation}
This equation is used to calculate the numbers in the second column of
Table~\ref{table4}.

\begin{table}[h!]
\caption{\label{table4}Values of $B/B_{0}$ at which
  $\varepsilon_{0}=-m_{e}$ according to the Dirac equation and nuclei
  become supercritical without (column 2,3) and with (column 4)
  taking screening into account.}
\begin{center}
\begin{tabular}{||c||c|c|c||}
\hline
$Z_{cr}$   & Eq. (\ref{eq:ORS32}) & Numerical results   &  Numerical results \\
${~}$ &       ${~}$                            &  without screening  &  with screening\\
\hline
90    &  118   &  116   &  122    \\
85    &  157   &  154   &  164    \\
80    &  213   &  210   &  229    \\
75    &  301   &  297   &  335    \\
70    &  444   &  438   &  527    \\
65    &  689   &  681   &  923    \\
60    &  1144  &  1133  &  1964   \\
55    &  2068  &  2053  &  6830   \\
54    &  2357  &  2340  &  10172  \\
53    &  2699  &  2681  &  17012  \\
52    &  3107  &  3087  &  35135  \\
51    &  3594  &  3572  &  $1.20\cdot 10^{5}$  \\
50    &  4181  &  4157  &  $1.14\cdot 10^{7}$  \\
45    &  9826  &  9787  &  ---  \\
40    &  28478 &  28408 &  ---  \\
35    &$1.12\cdot 10^{5}$& $1.12\cdot 10^{5}$  & ---  \\
30    &$6.99\cdot 10^{5}$& $6.98\cdot 10^{5}$  & ---  \\
25    &$9.27\cdot 10^{6}$& $9.27\cdot 10^{6}$  & ---  \\
\hline
\end{tabular}
\end{center}
\end{table}

\begin{figure}[ht!]
  \centering
  \includegraphics[width=0.9\textwidth]{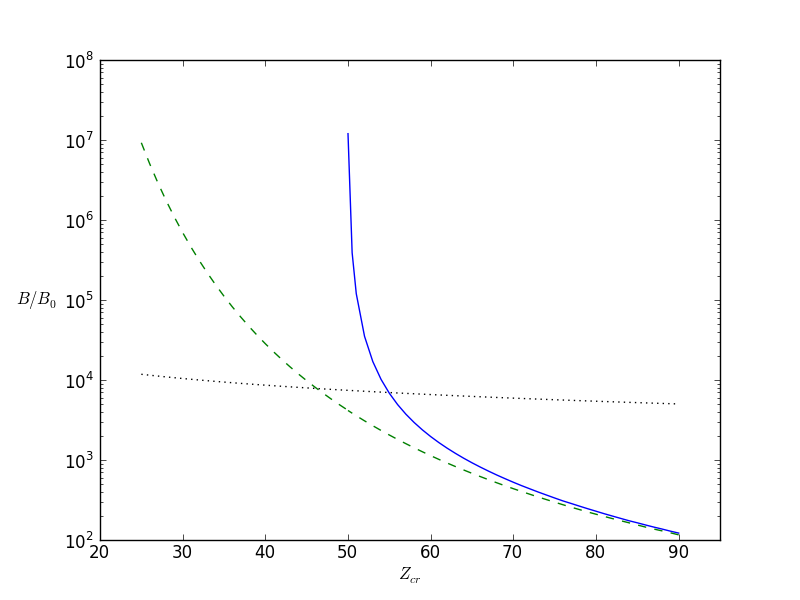}
  \caption{The values of $B^{Z}_{cr}$: a) without screening according
    to (\ref{eq:ORS32}), dashed (green) line; b) numerical results
    with screening, solid (blue) line. The dotted (black) line corresponds
    to the field at which $a_{H}$ becomes smaller than the size of the
    nucleus.}
  \label{fig:Bcr}
\end{figure}

From Table \ref{table4} we see that with the account of screening only the atoms with $Z \ga 52$ become
supercritical at the values of $B/B_0$ shown in the fourth column.
Because of screening a larger $B$ is needed for a nucleus to become
supercritical and the nuclei with $Z < 52$ never reach
supercriticality. This phenomenon is illustrated in Fig. \ref{fig:Bcr}.

From Tables \ref{tab:1}, \ref{tab:3}, and \ref{table4} we see that (\ref{9}) is very good in
describing the dependence of the energy on the magnetic field; at
least a numerical integration produces almost identical results.
In Table \ref{table6} we demonstrate several cases where the accuracy of
(\ref{9}) is not that good. It happens at low $B/B_0$ since
the matching condition $B > 4m_e^2/(e(Ze^{2})^2)$ fails and when
$\varepsilon_{0}$ is relativistic. However, $B$ should not be too low
to make the adiabaticity condition $a_B \gg
a_H$, or $B \gg (Ze^2)^2 m_e^2/e$ applicable.
\begin{table}[h!]
\caption{\label{table6}Values of $\varepsilon_{0}/m_{e}$ at $B/B_0=5$.}
\begin{center}
\begin{tabular}{||c||c|c||}
\hline
$Z$   & Eq. (\ref{9})        & Numerical results   \\
${~}$ &       (Dirac)        &  (Dirac)            \\
\hline
90  &0.2050     &0.2512         \\
80  &0.3096     &0.3539         \\
70  &0.4139     &0.4542         \\
60  &0.5171     &0.5516         \\
50  &0.6185     &0.6454         \\
40  &0.7165     &0.7349         \\
30  &0.8086     &0.8188         \\
20  &0.8914     &0.8952         \\
10  &0.9596     &0.9601         \\
 1  &0.998745   &0.998745       \\
\hline
\end{tabular}
\end{center}
\end{table}

Textbooks \cite{greiner} contain detailed consideration of the phenomenon of critical charge.

\section{Conclusions}
A magnetic field plays a double role in the critical charge
phenomenon. By squeezing the electron wave function and putting it in the
domain of a stronger Coulomb potential it diminishes
the value of the critical charge substantially \cite{3}. However, for
nuclei with $Z < 52$ to become critical such a strong $B$ is needed
that the screening of the Coulomb potential occurs and acts in the
opposite direction: the electron ground state energy freezes and the
nucleus remains subcritical in spite of growing $B$.

We are grateful to O. Babelon, V.L. Chernyak, V.S. Popov, A.I. Rez,
V.B. Semikoz, A.E. Shabad and S. Teber for useful discussions and
remarks. S.G. and M.V. were partially supported by the grant RFBR
11-02-00441 and by the grant of Russian Federation Government No.
11.G34.31.0047.

\end{document}